\documentclass[fdp,a4paper,fleqn%
]{w-art}
\usepackage{times,cite}

\usepackage[]{graphicx} 
\newcommand {\SO}{\mathop{\rm SO}}

\newcommand {\beq} {\begin{equation}}
\newcommand {\eeq} {\end{equation}}
\newcommand {\beqa}{\begin{eqnarray}}
\newcommand {\eeqa}{\end{eqnarray}}

\newcommand {\tr}{{\rm tr\,}}
\newcommand {\Tr}{\mbox{Tr\,}}

\newcommand {\1}{{\bf 1}}

\begin{document}
\DOIsuffix{theDOIsuffix}
\Volume{55}
\Month{01}
\Year{2007}
\pagespan{1}{}



\title[Recent developments in the type IIB matrix model]{Recent developments in the type IIB matrix model}


\author[J. Nishimura]{Jun Nishimura\inst{1,2,}%
  \footnote{E-mail:~\textsf{jnishi@post.kek.jp}}}
\address[\inst{1}]{Theory Center, 
High Energy Accelerator Research Organization (KEK),\\
1-1 Oho, Tsukuba, Ibaraki 305-0801, Japan}
\address[\inst{2}]{Department of Particle and Nuclear Physics,
Graduate University for Advanced Studies (SOKENDAI),\\
1-1 Oho, Tsukuba, Ibaraki 305-0801, Japan}
\begin{abstract}
We review recent developments in 
the type IIB matrix model, which
was conjectured to be 
a nonperturbative formulation of superstring theory.
In the first part we review the recent results 
for the Euclidean model, which suggest that SO(10) 
symmetry is spontaneously broken.
In the second part we review the recent results
for the Lorentzian model.
In particular, we discuss Monte Carlo results,
which suggest that (3+1)-dimensional expanding universe
emerges dynamically. We also discuss some results
suggesting the emergence of exponential expansion and the power-law
expansion at later times.
The behaviors at much later times are studied by
the classical equation of motion. We discuss 
a solution representing 3d expanding space, which suggests
a possible solution to the cosmological constant problem.
\end{abstract}
\maketitle                   





\section{Introduction}

Particle physics and cosmology have both entered a
difficult and yet interesting era.
We have successful phenomenological models supported by experiments
and observations, but we still lack fundamental understanding of these
models from a more microscopic point of view. 
On the particle physics side, we have the Standard Model,
which has been established, in particular, 
by the recent discovery of the Higgs particle
at LHC. But we do not have clear understanding for the origin 
of the Higgs particle, the matter fermions (the number of 
their generations, in particular) and the gauge group.
On the cosmology side, we have the standard scenario of Inflation 
and the Big Bang, which have been strongly supported by the
WMAP and PLANCK data for the cosmic microwave background,
the theory for structure formation of galaxies, nucleosyntheis and so on.
But we do not have clear understanding for the origin 
of the scalar field ``inflaton'', its potential
and its initial condition.
Furthermore,
we have serious naturalness problems in both particle physics and
cosmology, which are presumably related to the lack of 
fundamental understanding of the phenomenological models.
On the particle physics side, the hierarchy between the electroweak
scale and the Planck scale is unnatural, whereas on the cosmology side,
the accelerated expansion observed today suggests 
an unnaturally small but finite cosmological constant.
It is widely believed that these problems can be solved
by a fundamental theory that describes quantum gravity,
and superstring theory has been studied
as a promising candidate for such a theory.

The most fundamental issue in superstring theory concerns the
reason why the dimensionality of our space-time appears to be four
instead of ten as required from consistency of the theory.
A conventional approach towards this issue
is to compactify some of the dimensions
leading to infinitely many consistent vacua, which are perturbatively
stable. Each of them has different space-time dimensionality,
different gauge symmetry and different matter contents.
One can then hope to find a vacuum which explains what we observe now.
The discovery of D-brane led to many interesting new ideas
such as intersecting D-branes, D-brane Inflation and so on,
which enriched both string phenomenology and string cosmology.
However, an unavoidable feature of this conventional approach is that one 
obtains too many models and hence it is extremely hard to make predictions.
In this regard we should not forget that all these perspectives
are obtained from mostly perturbative studies of superstring theory,
including at most the nonperturbative effects represented by the
existence of D-branes.
Therefore, a totally new perspective might appear if one studies 
superstring theory in a completely nonperturbative framework
analogous to lattice gauge theory in the case of QCD.
Let us recall that nonperturbative aspects of QCD such as
confinement of quarks as well as the hadron mass spectrum can never be
understood from perturbation theory.

The type IIB matrix model was proposed as a nonperturbative
formulation of superstring theory \cite{IKKT}.
It is a theory that is expected to define superstring theory
beyond perturbative expansion just as lattice gauge theory
does so in the case of QCD.
The connection to perturbative formulations of superstring theory
can be seen manifestly by considering type IIB superstring theory 
in ten dimensions in the worldsheet formalism \cite{IKKT}
or in the light-cone string field formalism \cite{Fukuma:1997en}.
The model can be regarded as a natural extension \cite{AIKKT}
of the ``one-matrix model'',
which is established as a nonperturbative formulation of non-critical
strings \cite{Brezin:1990rb,Douglas:1989ve,Gross:1989vs}, 
where string worldsheets appear as Feynman 
diagrams in the matrix model and the large-$N$ limit can be taken
in such a way that diagrams with all different genera contribute.
Despite its manifest connection to type IIB superstring theory
within perturbation theory, the type IIB matrix model is expected
to provide the unique theory underlying the web of dualities
among various types of superstring theory.
For this to be true,
other types of superstring theory should be represented as
perturbative vacua of the type IIB matrix model.

The type IIB matrix model is given by the action 
$S=S_{\rm b}+S_{\rm f}$, where
\begin{eqnarray}
  S_{\rm b} &=& - \frac{1}{4 g^2}  \Tr 
[ A_\mu, A_\nu ] [ A^\mu, A^\nu ]  \ , 
\label{eq:sb} \\
  S_{\rm f} &=& - \frac{1}{2 g^2} 
\Tr\left(\Psi_\alpha ({\cal C} \Gamma^\mu)_{\alpha\beta}
    [ A_\mu, \Psi_\beta ] \right) \ . 
\label{eq:sf}
\end{eqnarray}
Here $A_\mu$ ($\mu = 1,\ldots,10$) are 
traceless $N\times N$ Hermitian matrices, 
whereas $\Psi_\alpha$ ($\alpha = 1,\ldots,16$) are 
traceless $N\times N$ matrices with Grassmannian entries. 
The model has an $\SO(9,1)$ symmetry, 
under which $A_\mu$ and $\Psi_\alpha$ transform as a vector
and a Majorana-Weyl spinor, respectively.
The Lorentz indices $\mu$ and $\nu$ in (\ref{eq:sb}) and (\ref{eq:sf})
are raised and lowered by the metric $\eta = {\rm diag}(-1,1,\ldots ,1)$.

Until quite recently,
the type IIB matrix model was studied 
after making a ``Wick rotation'',
which amounts to replacing the Hermitian matrix $A_0$
by $A_0 = i A_{10}$,
and treating the Hermitian matrix $A_{10}$ 
on equal footing as the matrices $A_i$ ($i=1,\ldots ,9$)
in the spatial directions.
The Euclidean model obtained in this way has 
manifest SO(10) symmetry, and it
is well defined as Monte Carlo studies
with small matrices demonstrate \cite{Krauth:1998xh}.
In fact the partition function was proven to be finite
for arbitrary matrix size \cite{AW}.
In ref.~\cite{AIKKT},
perturbative expansion around
the diagonal configurations $A_\mu = {\rm diag}(x_{1\mu} , \ldots ,
x_{N\mu} )$ 
was studied and
the low-energy effective
theory for the diagonal elements
was discussed.
In particular, it was speculated that configurations with the $N$ points
$\{ \vec{x}_{i} ; i=1,\ldots , N\}$ distributed on a four-dimensional
hypersurface in ten-dimensional Euclidean space
may be favored due to some nontrivial interactions
in the low-energy effective theory.
If that really happens, it implies that 
the SO(10) symmetry is spontaneously
broken down to SO(4) and 
that four-dimensional space-time is
generated dynamically.

The recent developments in the type IIB matrix model we would like
to discuss are the following.
First in the Euclidean model, it was found that 
SSB indeed occurs, but the SO(10) symmetry is 
broken down to SO(3) \cite{Nishimura:2011xy}.
The extent of space-time in the extended direction
and that in the shrunken direction
are calculated, and their ratio is found to be around five
in the large-$N$ limit.
Interpretation of these results is unclear, though,
since the Wick rotation is not justifiable unlike in ordinary quantum
field theory.

On the other hand, it was found that
the Lorentzian model can be made well-defined by introducing infrared
cutoffs and removing them in the large-$N$ limit \cite{KNT1}.
Real-time evolution can be extracted from matrix configurations
that dominate the path integral.
It was shown that expanding 
three-dimensional space appears after a critical time \cite{KNT1},
and the possibility of observing Inflation and the Big Bang has
been discussed \cite{Ito:2013qga,Ito:2013ywa}.
The behavior at much later times has been studied by 
solving the classical equation of motion \cite{Kim:2011ts}, and 
a natural solution to the cosmological constant problem has been 
suggested \cite{KNT2}.
Also the realization of the Standard Model in the type IIB matrix model
has been 
discussed \cite{Chatzistavrakidis:2011gs,Aoki:2014cya,Steinacker:2014fja}
by applying the idea of intersecting branes.
The main message we would like to convey is that
the Lorentzian version of type IIB matrix model seems to be indeed
the correct nonperturbative formulation of superstring theory, which
describes our Universe.

The rest of this article is organized as follows.
In section \ref{sec:Euclidean} we discuss the recent results
obtained in the Euclidean type IIB matrix model.
In section \ref{sec:Lorentzian} we discuss the recent results
obtained in the Lorentzian type IIB matrix model.
Section \ref{sec:summary} is devoted to a summary and future prospects.

\section{Euclidean type IIB matrix model}
\label{sec:Euclidean}

In this section we discuss the recent results obtained
in the Euclidean type IIB matrix model.

\begin{figure}
\includegraphics[width=68mm]{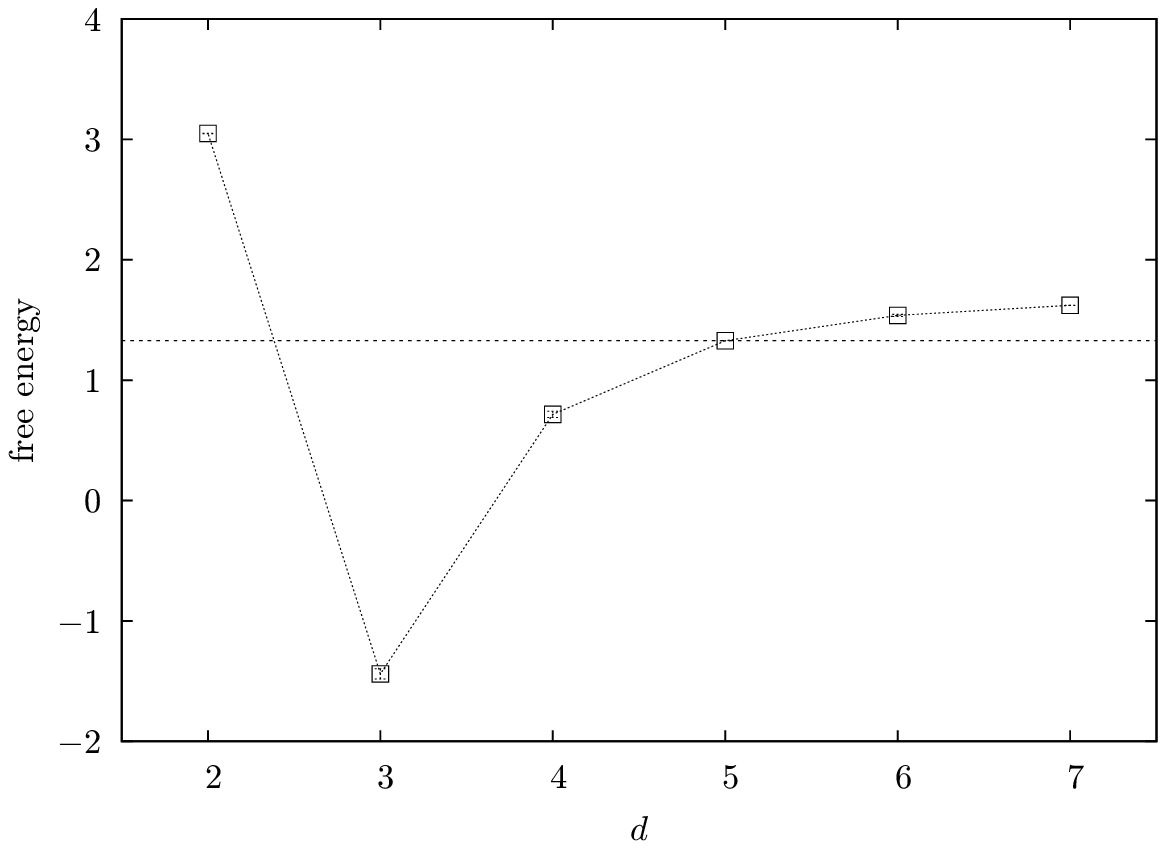}~a)
\hfil
\includegraphics[width=68mm]{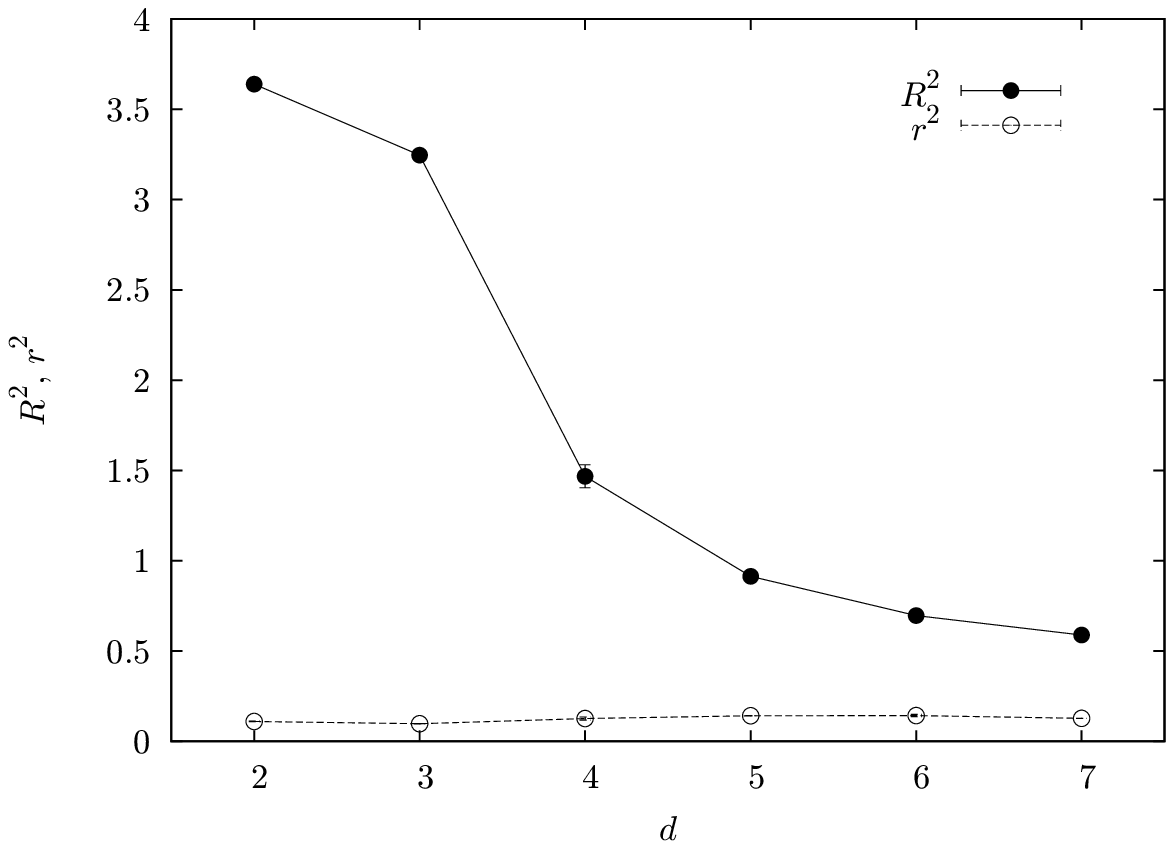}~b)
\caption{a) The free energy for the SO($d$) symmetric vacuum
is plotted against $d$. 
The horizontal
line represents the value $f=\log 8 - \frac{3}{4} = 1.32944\ldots$
corresponding to the prediction in ref.~\cite{Krauth:1998xh}.
b) The extent of space-time $R^2$ and $r^2$ 
in the extended
and shrunken directions, respectively, 
are plotted against $d$.
}
\label{fig:free-summary}
\end{figure}

In Fig.~\ref{fig:free-summary} we show the 
most recent results \cite{Nishimura:2011xy}
based on the Gaussian expansion 
method \cite{Nishimura:2001sx} in the large-$N$ limit.
On the left we plot the free energy
for the SO($d$) symmetric vacuum.
We find that $d=3$ gives the minimum free energy,
which implies that SO(10) symmetry is broken down spontaneously
to SO(3).
On the right we plot the extent of space-time 
in the extended directions (filled circles)
and that in the shrunken directions (open circles).
We find that 
the former increases as $d$ decreases, whereas
the latter is almost independent of $d$.

The Gaussian expansion method has been applied also to 
a 6d version of the type IIB matrix model, which can be obtained
by dimensionally reducing 6d super Yang-Mills (SYM) theory 
to a point \cite{Aoyama:2010ry},
and it was found that the SO(6) symmetry is spontaneously broken
down to SO(3).
The mechanism of SSB is demonstrated by Monte Carlo studies 
in this case \cite{Anagnostopoulos:2013xga}.

In order to probe the SSB of SO($6$) rotational symmetry,
we studied the ``moment of inertia'' tensor
\begin{equation}
\label{2.9}
T_{\mu\nu}=\frac{1}{N} \Tr\left(A_\mu A_\nu\right)
\end{equation}
and its real positive eigenvalues
$\lambda_n$ ($n=1,\ldots,6$) ordered as
$\lambda_1\ge\lambda_2\ge\ldots\ge\lambda_6$.
The vacuum expectation values (VEVs) $\langle  \lambda_n \rangle$, 
taken {\it after}
the ordering for each configuration, play the role of order
parameters. If they turn out to be unequal in the large-$N$ limit, it
implies SSB of SO($6$). 
It has been speculated that the phase of the
fermion determinant induces the SSB \cite{NV,Anagnostopoulos:2001yb}. 
However, the effect of the phase is difficult to implement 
in Monte Carlo calculation due to the so-called sign problem.

\begin{figure}
\includegraphics[width=68mm]{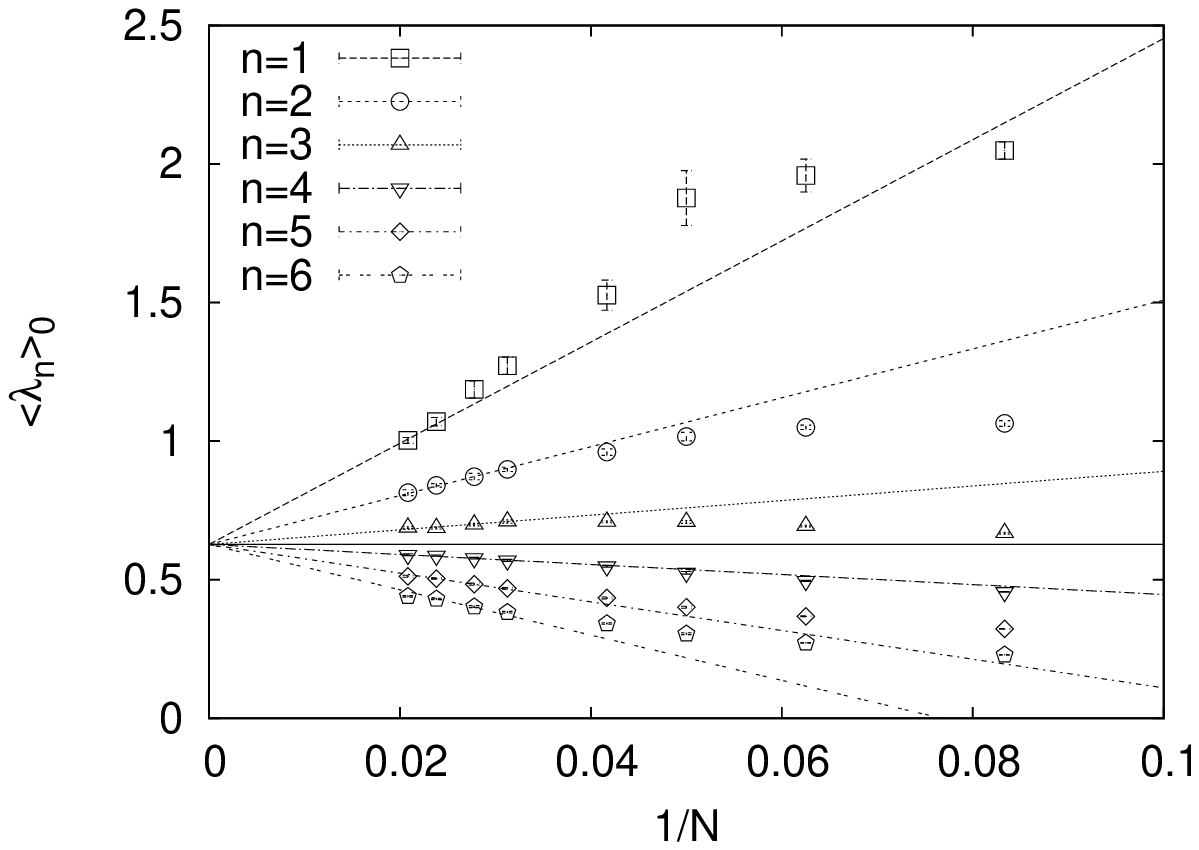}~a)
\hfil
\includegraphics[width=68mm]{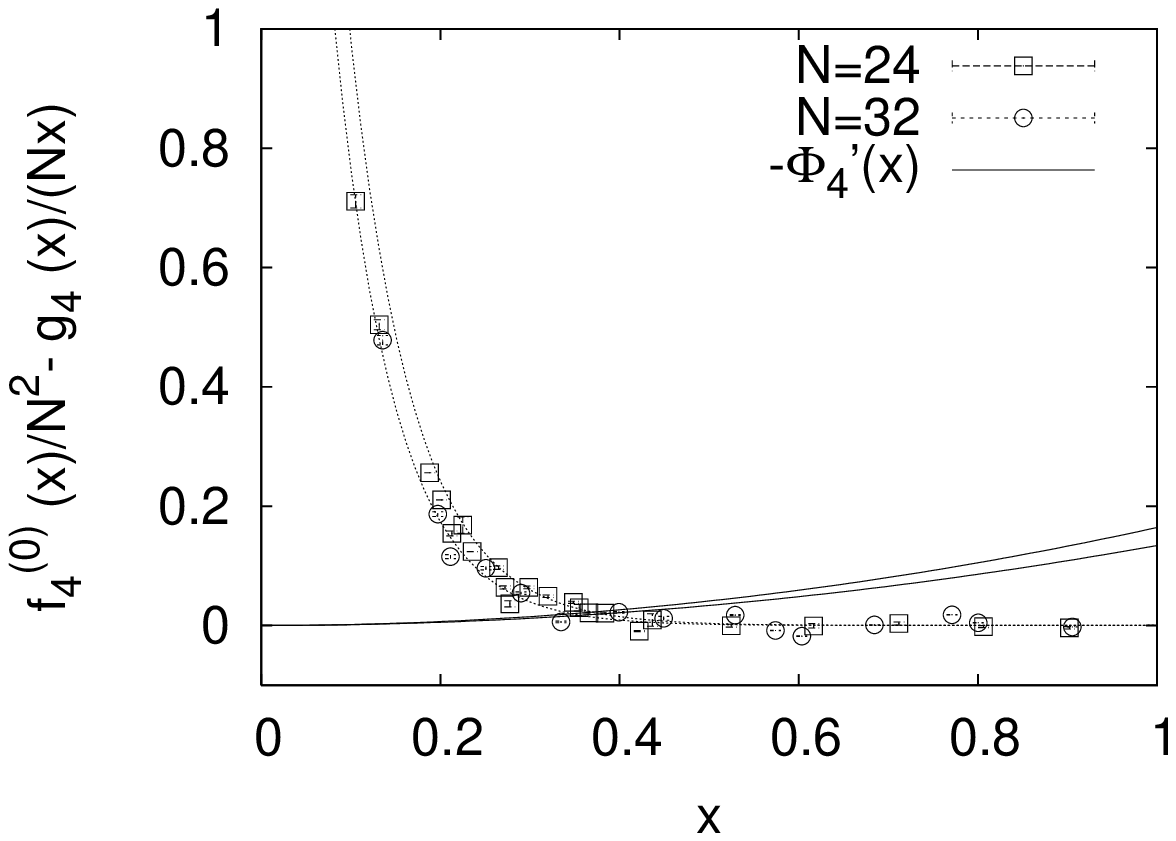}~b)
\caption{a) The eigenvalues 
$\langle \lambda_{n} \rangle_{0}$ for the phase-quenched model
are plotted against $1/N$. 
The solid line represents the value
$\ell^2 = 0.627$, which is predicted
by the Gaussian expansion method \cite{Aoyama:2010ry}.
The other lines represent the fits to the behavior 
$\langle \lambda_n \rangle_0 = \ell^2 + {\cal O}(1/N)$.
b) The Monte Carlo results obtained for the SO(3) symmetric vacuum.
The intersection $x\sim 0.35$ gives the extent of space-time 
in the shrunken directions, which turns out to be in good agreement
with the prediction (\ref{compare-GEM})
by the Gaussian expansion method.
}
\label{fig:MC-SO2345}
\end{figure}

In Fig.~\ref{fig:MC-SO2345}a we show 
the results for $\langle  \lambda_n \rangle_0$ obtained in a model,
which is obtained by simply omitting 
the phase of the fermion determinant \cite{Anagnostopoulos:2013xga}.
We find that all the eigenvalues converge to the same value 
($ \ell^2 \sim 0.627$) at large $N$.
In ref.~\cite{Anagnostopoulos:2013xga}
the effect of the phase has been studied by an analysis
based on the factorization 
method \cite{Anagnostopoulos:2001yb,Anagnostopoulos:2010ux}.
In Figs.~\ref{fig:MC-SO2345}b
we show the results obtained for 
the SO(3) symmetric vacuum.
See the original paper \cite{Anagnostopoulos:2013xga} for the details.
From the intersecting point, we can obtain the extent of space-time
in the shrunken directions.
We find that it is given by $x\sim 0.35$, which
should be compared with the value obtained 
by the Gaussian expansion method \cite{Aoyama:2010ry}
\begin{equation}
\label{compare-GEM}
x= \frac{r^2}{\ell^2} \sim \frac{0.223}{0.627} \sim 0.355 \ ,
\end{equation}
taking account of the chosen normalization. 
Thus we find that the two completely different methods give
consistent results, which supports the validity of both calculations.



After all, we consider that the problem was in the Euclideanization.
In quantum field theory, it can be fully justified as analytic
continuation. That's why we can use lattice gauge theory.
On the other hand, it is subtle in gravitating theory,
although it might be OK at the classical level.
For instance, in quantum gravity based on the dynamical triangulation
approach, it was found that problems with Euclidean gravity
can be overcome in Lorentzian gravity \cite{Ambjorn:2005qt}.
As another example, we quote Coleman's worm hole scenario for the
cosmological constant problem.
It was found recently \cite{Kawai:2011rj}
that a physical interpretation is possible
only by considering the Lorentzian version instead of the original
Euclidean version.
Moreover, Euclidean theory is useless for studying the real time
dynamics such as the expanding Universe.
All these considerations led us to consider the Lorentzian version
of the type IIB matrix model.

\section{Lorentzian type IIB matrix model}
\label{sec:Lorentzian}

In this section we discuss recent results obtained in 
the Lorentzian type IIB matrix model.

\subsection{Definition of Lorentzian type IIB matrix model}

We define the partition function of the Lorentzian model by \cite{KNT1} 
\beq
Z = \int d A \, d\Psi \, e^{i S} =
\int d A \,  e^{i S_{\rm b}} {\rm Pf}{\cal M}(A)
\ ,
\label{partition-fn-def}
\eeq
where the Pfaffian ${\rm Pf}{\cal M}(A)$ appears from integrating
out the fermionic matrices $\Psi_\alpha$.
Note that in the Euclidean model,
the Pfaffian is complex in general,
and its phase plays a crucial role in the 
SSB of SO(10) symmetry as we have seen in the previous section.
On the other hand, the Pfaffian in the Lorentzian model is \emph{real}.
Therefore, the mechanism of SSB that was identified
in the Euclidean model is absent in the Lorentzian model.

In the definition (\ref{partition-fn-def}),
we have replaced the ``Boltzmann weight'' $e^{-S}$
in the Euclidean model by $e^{iS}$.
This is theoretically motivated
from the connection
to the worldsheet theory \cite{IKKT}.
The partition function
(\ref{partition-fn-def})
can also be obtained formally
from pure ${\cal N}=1$ SYM theory
in $(9+1)$ dimensions by dimensional reduction.
Note, however, that
the expression (\ref{partition-fn-def}) is ill-defined and
requires appropriate regularization
in order
to make any sense out of it.
It turns out that the integration over $A_\mu$ is divergent,
and we need to introduce two constraints
\beqa
\frac{1}{N}\Tr (A_0)^2  &\le&  \kappa \frac{1}{N} \Tr (A_i)^2  \ ,
\label{T-constr} \\
\frac{1}{N} \Tr (A_i)^2   &\le&  L^2  \ .
\label{R-constr}
\eeqa
This is in striking contrast to the Euclidean model,
in which the partition function is shown to
be finite without any regularization \cite{Krauth:1998xh,AW}.

Note that $e^{iS_{\rm b}}$ in the partition function (\ref{partition-fn-def})
is a phase factor
just as in the path-integral formulation of quantum field theories
in Minkowski space.
However, we can circumvent the sign problem by integrating out the scale factor
of $A_\mu$, which essentially replaces the phase $e^{iS_{\rm b}}$ by
the constraint $S_{\rm b} \approx 0$. (Such a constraint is analogous
to the one that appeared in the model inspired by space-time uncertainty
principle \cite{Yoneya:1997gs}.)
Without loss of generality, we set $L=1$ in (\ref{R-constr}), 
and thus we arrive at the model \cite{KNT1}
\beq
Z =  \int dA \,
\delta \left(
\frac{1}{N}\Tr (F_{\mu\nu}F^{\mu\nu})  \right)
 {\rm Pf} {\cal M} (A)
\, \delta\left(\frac{1}{N}\Tr (A_i)^2 - 1 \right)
\theta\left(\kappa  - \frac{1}{N}\Tr (A_0)^2  \right)  \ ,
\label{our-model}
\eeq
where $\theta(x)$ is the Heaviside step function.
Since the Pfaffian ${\rm Pf} {\cal M}(A)$ is real
in the present Lorentzian case,
the model (\ref{our-model}) can be studied by Monte Carlo simulation
without the sign problem.\footnote{Strictly speaking, the Pfaffian
can flip its sign, but we find that the configurations with
positive Pfaffian dominate as $N$ is increased.
Hence, we just take the absolute value of the Pfaffian in actual simulation.} 
Note that this is usually not the case for quantum field theories in
Minkowski space.
\subsection{Expanding 3d out of 9d}
\label{sec:MClorentz}

In ref.~\cite{KNT1}
we performed Monte Carlo simulation of the model (\ref{our-model}).
In order to extract the ``time evolution'', we diagonalize $A_0$,
and define the eigenvectors $| t_a \rangle$ corresponding
to the eigenvalues $t_a$ of $A_0$ ($a=1 , \ldots , N$)
with the specific order $t_1 < \ldots < t_N$.
The spatial matrix in this basis $\langle t_{a} | A_i | t_{b} \rangle $
is not diagonal, but it turns out that the off-diagonal elements
decrease rapidly as one goes away from a diagonal element.
This motivates us to
define $n\times n$ matrices
$\bar{A}_i^{(ab)}(t) \equiv  \langle t_{\nu+a} | A_i | t_{\nu+b} \rangle $
with $1 \le a , b \le n$ and
$t= \frac{1}{n}\sum_{a=1}^{n} t_{\nu + a}$
for $\nu=0,\ldots , (N-n)$.
These matrices represent the state of the Universe
at fixed time $t$.
(This point of view can be justified in the large-$N$ limit,
in which more and more eigenvalues of $A_0$ 
appear around some value $t$
within a fixed interval $\delta t$.)
The block size $n$ should be large enough to include non-negligible 
off-diagonal elements.

\begin{figure}
\includegraphics[width=68mm]{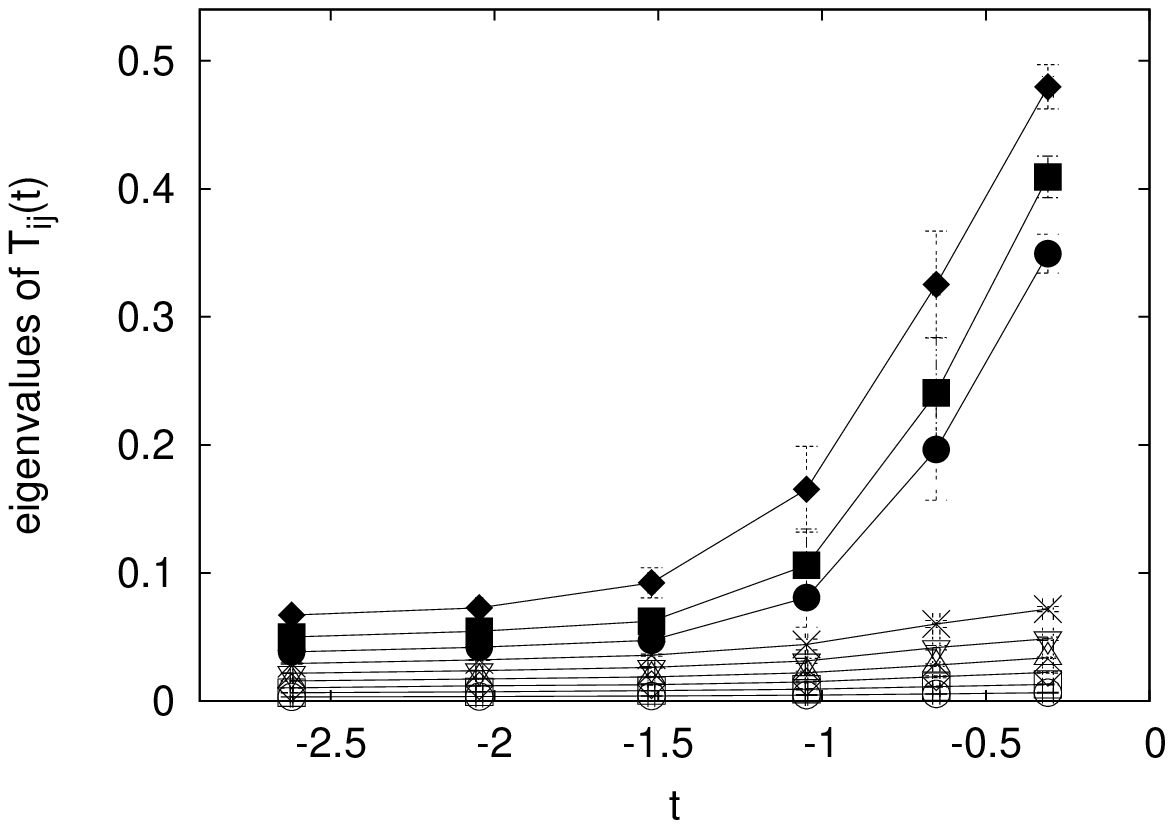}~a)
\hfil
\includegraphics[width=68mm]{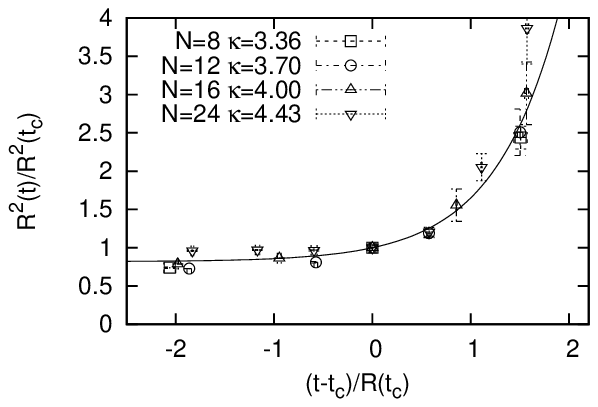}~b)
\caption{a) The 9 eigenvalues of $T_{ij}(t)$ with $N=16$ and $n=4$
are plotted as a function of $t$
for $\kappa=4.0$. After the critical time $t_{\rm c}$,
3 eigenvalues become larger,
suggesting that the SO(9) symmetry is spontaneously broken
down to SO(3).
b) The extent of space 
$R^2\left(t\right)/R^2\left(t_{c}\right)$
is plotted against $\left(t-t_{c}\right)/R\left(t_{c}\right)$ 
for the model (\ref{our-model})
with various $\kappa$ and $N$.
%
The solid line is a fit to the exponential behavior
$y=a+ (1-a) \exp (bx)$ with $a=0.82(1)$ and $b=1.5(2)$.
}
\label{fig:expo-LIKKT}
\end{figure}

Let us study the spontaneous breaking of the SO(9) symmetry.
As an order parameter,
we define the $9 \times 9$
(positive definite)
real symmetric tensor \cite{KNT1}
\beq
T_{ij}(t) = \frac{1}{n} \tr \Bigl\{
\bar{A}_i(t) \bar{A}_j(t) \Bigr\} \ ,
\label{Tij-def}
\eeq
which is an analog of (\ref{2.9})
in the Euclidean model.
The trace ``$\tr$'' used here is taken over the $n \times n$ matrices.
The 9 eigenvalues of $T_{ij}(t)$ are
plotted against $t$ in Fig.~\ref{fig:expo-LIKKT}a.
We find that 3 largest eigenvalues of
$T_{ij}(t)$ start to grow at the critical
time $t_{\rm c}$, which suggests that the SO(9)
symmetry is spontaneously broken down to
SO(3) after $t_{\rm c}$.

\subsection{Exponential/power-law expansion}

In Fig.~\ref{fig:expo-LIKKT}b
we plot the extent of space 
\begin{equation}
R(t)^2 \equiv \frac{1}{n} \tr \bar{A}_i(t)^2 
\label{eq:R(t)}
\end{equation}
presented in ref.~\cite{Ito:2013qga}.
(Note that $R(t)^2$ is given by the sum of 9 eigenvalues
of $T_{ij}(t)$ defined in (\ref{Tij-def}).)
We normalize dimensionful quantities by
$R(t_{\rm c})$, where $t_{\rm c}$ is the ``critical time''
at which the spatial SO(9) symmetry is spontaneously
broken down to SO(3).
In fact the obtained $R(t)$ can be nicely fitted 
with $y=f(x)\equiv a+ (1-a) \exp (bx)$,
where we have imposed $f(0)=1$, 
which follows from the chosen normalization.
This implies that three spatial directions actually start to
expand exponentially, 
which may be interpreted as the beginning
of Inflation.


In order to confirm the exponential behavior for 
a longer time period,
we need to increase the matrix size further, which makes 
the simulation too time-consuming.
In ref.~\cite{Ito:2013ywa}
we considered, instead, a simplified model that describes the behavior
at early times.
To motivate the model, let us 
decompose the fermionic action (\ref{eq:sf}) into
two terms as
\begin{equation}
S_{{\rm f}} 
 \propto 
{\rm Tr}\left(\Psi_{\alpha}
\left(C\Gamma^{0}\right)_{\alpha\beta}
\left[A_{0},\Psi_{\beta}\right]\right)+
{\rm Tr}\left(\Psi_{\alpha}\left(C\Gamma^{i}\right)_{\alpha\beta}
\left[A_{i},\Psi_{\beta}\right]\right) \ .
\label{eq:fermionic action}
\end{equation}
Due to the expanding behavior of the universe,
the elements of the spatial matrices $A_i$ become very large
at late times.
At early times, on the other hand,
it is expected that the first term 
in (\ref{eq:fermionic action}) is more important,
so we simply omit the second term 
in \eqref{eq:fermionic action} as a simplification. 
Integrating out the fermionic matrices\footnote{Strictly speaking,
there are zero modes corresponding to $\Psi_\alpha$
satisfying $[A_0 , \Psi_\alpha] = 0$, which we simply neglect.}, 
we obtain the Pfaffian, which is now given by 
\begin{equation}
{\rm Pf}\mathcal{M}\left(A\right)=\Delta^{2(d-1)} \ , 
\label{eq:Pf early}
\end{equation}
where $\Delta \equiv \prod_{i>j}\left(\alpha_{i}-\alpha_{j}\right)$ 
is the van der Monde determinant
and we have written down the general results for dimensionally
reduced SYM models with $d$ spatial dimensions 
($d=9$ in the case of type IIB matrix model). 
The Pfaffian \eqref{eq:Pf early} obtained here causes
a repulsive force between all the pairs of eigenvalues of $A_{0}$,
which cancels the attractive force arising from
the fluctuation of the bosonic matrices at the one-loop level.
Due to this cancellation,
the eigenvalues of $A_{0}$ can extend to infinity, which 
necessitates the cutoff (\ref{T-constr})
in the temporal direction. 

\begin{figure}
\includegraphics[width=68mm]{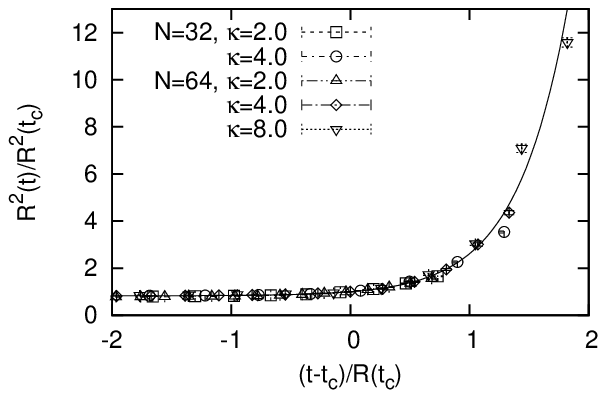}~a)
\hfil
\includegraphics[width=68mm]{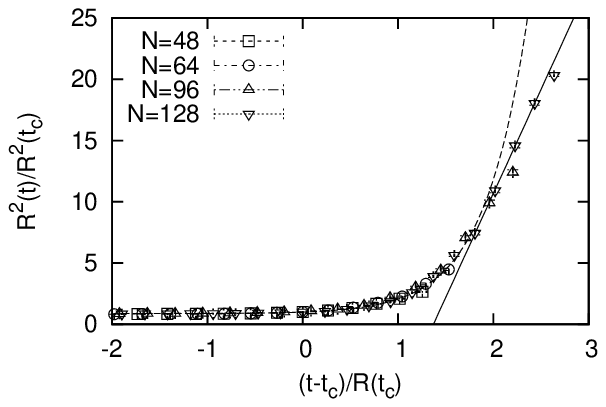}~b)
\caption{a) The extent of space 
$R^{2}\left(t\right)/R^{2}\left(t_{\rm c}\right)$
is plotted against 
$\left(t-t_{\rm c}\right)/R\left(t_{\rm c}\right)$ 
for the simplified model for early times 
in the $d=5$ case with various $\kappa$ and $N$.
The solid line is a fit to the exponential behavior
$y=a+ (1-a) \exp (bx)$ with $a=0.83(1)$ and $b=2.3(1)$.
b) The extent of space 
$R^{2}\left(t\right)/R^{2}\left(t_{\rm c}\right)$
is plotted against 
$\left(t-t_{\rm c}\right)/R\left(t_{\rm c}\right)$ 
for the $d=5$ quenched model with various $N$,
which is considered as a simplified model for late times.
The dashed line represents a fit
$y=a+\left(1-a\right)e^{bx}$
to the early time behavior ($a=0.870(3)$, $2.21(3)$), 
whereas the solid line represents
a fit $y=cx+d$ to the late time behavior ($c=17.0(1)$, $d=-23.3(3)$).
}
\label{fig:bosonic}
\end{figure}

The simplified model for early times
with the Pfaffian replaced by (\ref{eq:Pf early})
can be simulated with much less efforts.
In ref.~\cite{Ito:2013ywa} we studied the $d=5$ model,
in which the rotational SO(5) symmetry is broken down 
to SO(3) at some critical time $t_{\rm c}$
analogously to the $d=9$ model.
In Fig.~\ref{fig:bosonic}a we plot the extent of space
\eqref{eq:R(t)}
as a function of $t$ for various $N$ and $\kappa$. 
This confirms the exponentially expanding behavior
in the simplified model,
which suggests that 
the first term of 
the fermionic action \eqref{eq:fermionic action}
is indeed important for the space to expand exponentially.

At late times, 
the second term in the fermionic
action (\ref{eq:fermionic action}) 
becomes more important, and it is expected that 
the repulsive force represented by (\ref{eq:Pf early})
is no more effective. In order to mimic such a situation,
we considered a quenched model obtained by omitting the
fermionic matrices completely \cite{Ito:2013qga}.
In this model,
since the eigenvalues of $A_{0}$ attract each other,
we do not need to introduce the cutoff (\ref{T-constr})
in the temporal direction.
The extent of the eigenvalue distribution 
increases with $N$, however, 
and one can take both the continuum and infinite-volume limits.
The breaking of SO(5) symmetry (for $d=5$)
down to SO(3) is observed 
after a critical time $t_{\rm c}$
for sufficiently large matrix size $N$.

In Fig.~\ref{fig:bosonic}b we plot the extent of space
\eqref{eq:R(t)} for the $d=5$ quenched model.
The exponential behavior is observed for some period after the
critical time, but it changes into a linear behavior
$R^{2}(t) \sim t$ meaning that $R (t) \sim t^{1/2}$,
which agrees with the expanding behavior
of the 
Friedmann-Robertson-Walker (FRW) 
universe in the radiation dominated era.
It would be interesting to confirm 
the transition from the exponential behavior
to the power-law behavior directly 
in the original model (\ref{our-model}).
In particular, this will tell us the value of E-folding,
which is determined \emph{dynamically}
in the Lorentzian type IIB matrix model.

\subsection{Time-evolution at much later times}

While the behaviors at much later times
are difficult to study by
direct Monte Carlo methods,
the classical equation of motion is expected to become more and more
valid at later times
since the value of the action increases 
with the cosmic expansion \cite{Kim:2011ts}.
There are actually
many classical solutions,
which is reminiscent of the fact
that superstring theory possesses infinitely many vacua 
that are perturbatively stable.
However, unlike in perturbative superstring theory,
we have the possibility to pick up the unique solution
that describes our universe because we have
a well-defined partition function.
In particular, we find a classical solution with an expanding
behavior that can naturally solve the cosmological constant 
problem \cite{KNT2}.
(See refs.~\cite{Steinacker:2011wb,Chatzistavrakidis:2011su}
for other works on the classical solutions in the type IIB matrix model.)

When we search for classical solutions
in the Lorentzian model,
it is important to take account of
the two cutoffs that 
had to be introduced in order to make the model well-defined.
Since the inequalities 
(\ref{T-constr}) and (\ref{R-constr})
are actually saturated
as is also seen by Monte Carlo simulation \cite{KNT1},
we search for stationary points of
the bosonic action $S_{\rm b}$ 
for fixed $\frac{1}{N} \Tr (A_0)^2$ and $\frac{1}{N} \Tr (A_i)^2$.
Then we have to extremize the function
\begin{align}
\tilde{S}=\Tr
\left( - \frac{1}{4}[A_{\mu},A_{\nu}][A^{\mu},A^{\nu}]
+\frac{\tilde{\lambda}}{2}(A_0^2-\kappa L^2)
-\frac{\lambda}{2}(A_i^2-L^2)\right) \ ,
\label{tildeS}
\end{align}
where $\lambda$ and $\tilde{\lambda}$ are the Lagrange multipliers.
Differentiating (\ref{tildeS}) with respect to $A_0$ and $A_i$, 
we obtain
\begin{align}
-[A_0,[A_0, A_i]]+[A_j,[A_j, A_i]] - \lambda A_i &= 0 \ , \label{cl-eq1} \\
[A_j,[A_j, A_0]]  - \tilde{\lambda} A_0 &= 0 \ ,
\label{cl-eq2}
\end{align}
respectively. 


A general prescription to solve 
the equations of motion (\ref{cl-eq1}) and (\ref{cl-eq2})
is given as follows \cite{KNT2}.
Let us first define a sequence of commutation relations
\begin{align}
[A_i,A_j]=iC_{ij} \ , \quad
[A_i,C_{jk}]=iD_{ijk}\ , \quad
[A_0,A_i]=iE_i \ , \label{E}\\
[A_0,E_i]=iF_i \ , \quad
[A_i,E_j]=iG_{ij} \ , \  \ldots  \ , \label{G}
\end{align}
where $1 \leq i,j,k \leq 9$ and the symbols
on the right-hand side represent 
Hermitian operators newly defined. 
Then we determine the relationship 
among $A_0$, $A_i$, $C_{ij}$, $D_{ijk}$, $E_i$, $F_i$, $G_{ij},\ldots$
so that
the equations of motion (\ref{cl-eq1}) and (\ref{cl-eq2})
and the Jacobi identities are satisfied. 
We obtain a Lie algebra in this way.
Considering that all the operators are Hermitian,
each unitary representation of the Lie algebra
gives a classical solution.



As an example of SO(4) symmetric solution, we consider \cite{KNT2}
\begin{align}
A_0= b T_0 \otimes \1_k \ , \quad \quad
A_i= \alpha b T_1 \otimes M_i \quad (i=1,2,3,4) \ ,
\label{new solution}
\end{align}
where
$T_0$ and $T_1$ are the two generators of 
the ${\rm SU}(1,1)$ algebra
\begin{align}
[T_0, T_1] = i T_2 \ , 
\quad [T_2, T_0] = i T_1 \ ,
\quad  [T_1, T_2] = -i T_0 \ .
\label{SU(1,1)}
\end{align}
$M_i$ are $k\times k$ diagonal matrices defined by 
\begin{align}
M_i = \mbox{diag}(n^{(1)}_{i},n^{(2)}_{i},\ldots,n^{(k)}_{i})  \ ,
\quad \quad
|n^{(I)}|=1\;\;(I=1,\ldots,k) \ ,
\label{condition for r}
\end{align}
and $\1_k$ is the $k\times k$ unit matrix. 
This is a solution of (\ref{cl-eq1}) and (\ref{cl-eq2})
for $\lambda=-b^2$ and $\tilde{\lambda}=-\alpha^2 b^2$,
and it represents $(3+1)$-dimensional space-time
with $R\times S^3$ geometry.

Let us discuss the cosmological implications
of this solution.
As an irreducible unitary representation of the ${\rm SU}(1,1)$ 
algebra, we consider
the primary unitary series 
representation,
in which the matrix elements of the generators are given as
\begin{align}
(T_0)_{mn}&= n \delta_{mn}\ , \nonumber\\
(T_1)_{mn}&=-\frac{i}{2}(n-i\rho)\delta_{m,n+1}
+\frac{i}{2}(n+i\rho)\delta_{m,n-1} \ , \nonumber\\
(T_2)_{mn}&=-\frac{1}{2}(n-i \rho)\delta_{m,n+1}
-\frac{1}{2}(n+i \rho)\delta_{m,n-1} \ ,
\label{matrix elements of SU(1,1)}
\end{align}
where $m,n \in {\bf Z}$.
In this case, $A_1$ has a tri-diagonal structure.
Therefore, we extract $3\times 3$ submatrices
$\bar{A}_0(n)$ and $\bar{A}_1(n)$.
Then we find that the extent of space $R(n)$ at a discrete time $n$ 
becomes
\begin{align}
R(n)=\sqrt{\frac{\alpha^2 b^2}{3}\left(n^2+\rho^2+\frac{1}{4}\right)} \ .
\end{align}
Let us take the continuum limit.
We define the continuum time by $t=n b$ and take 
the $b\rightarrow 0$ limit.
We also take the $\rho\rightarrow \infty$ limit
at the same time so that $t_0\equiv \rho b$ is kept fixed.
Then $R(t)$ is given by
\begin{align}
R(t)=\sqrt{\frac{\alpha^2}{3}(t^2 + t_0^2)} \ .
\label{RtPUSR}
\end{align}

\begin{figure}
\includegraphics[width=68mm]{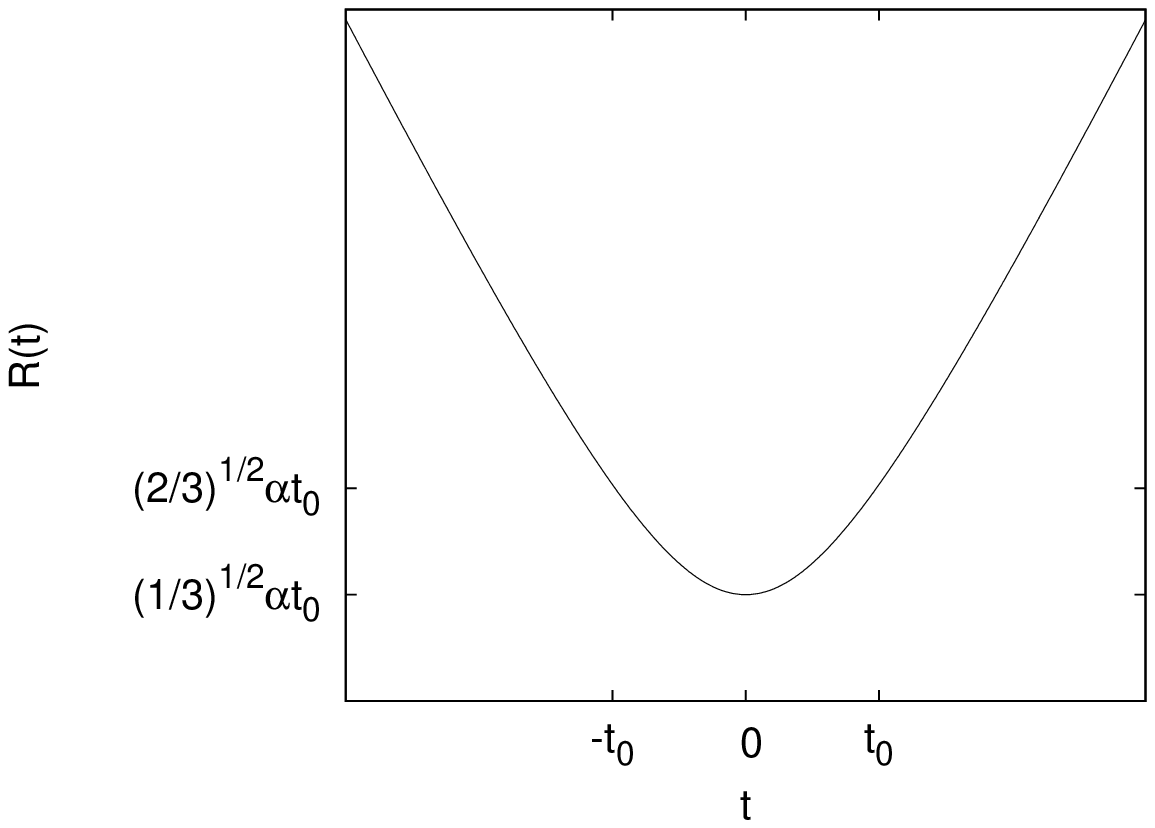}~a)
\hfil
\includegraphics[width=68mm]{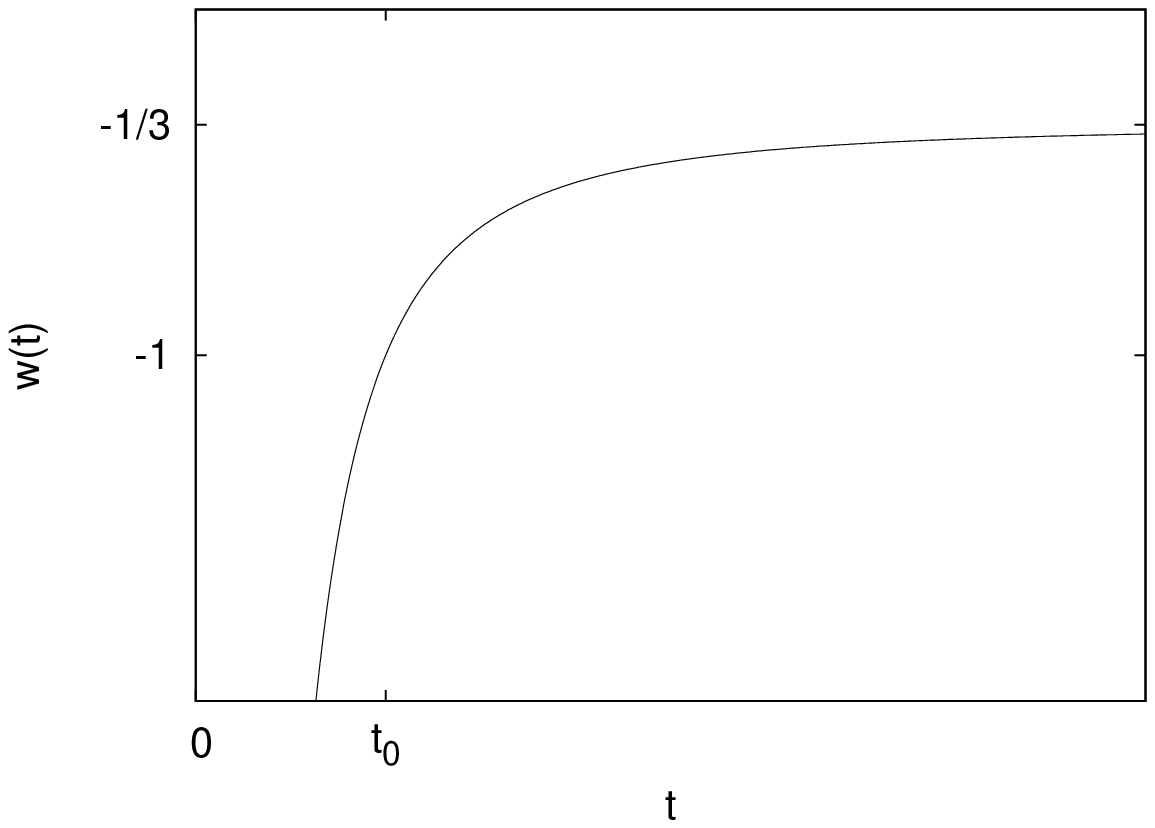}~b)
\caption{a) The time dependence of the scale factor $R(t)$
for the ${\rm SU}(1,1)$ solution with the 
primary unitary series 
representation.
b) The time dependence of the parameter $w$ 
for the same solution is shown.
}
\label{fig:su11pri_rt-wt}
\end{figure}

Here we naively identify $R(t)$ with the scale factor
of the FRW universe.
Then, we obtain the Hubble parameter $H$ and the parameter $w$ as
\begin{align}
H \equiv \frac{\dot{R}}{R} =
\frac{\alpha}{\sqrt{3}R^2}\sqrt{R^2 - \frac{\alpha^2t_{0}^2}{3}}
 \ , 
\quad
w \equiv - \frac{2R}{3} \frac{d \ln H}{dR} -1 =
- \frac{2t_{0}^2}{3t^2}-\frac{1}{3} \ ,
\end{align}
which are plotted against $t$ 
in Figs.~\ref{fig:su11pri_rt-wt}a and b, respectively.
We find that $w$ converges to $-\frac{1}{3}$ 
as $t\rightarrow\infty$, which corresponds to
the expansion of universe with a constant velocity.

If we identify $t_0$ with
the present time, the present value of $w$ is $-1$. 
This value of $w$ corresponds to
the cosmological constant, 
which explains the present accelerating expansion of the universe. 
Moreover, the corresponding cosmological constant
becomes of the order of $(1/t_0)^4$,
which suggests a possible solution to the cosmological constant problem.
As we mentioned above, $w$ increases with time and approaches $-\frac{1}{3}$.
This means that the cosmological constant actually vanishes in the future. 

\section{Summary and future prospects}
\label{sec:summary}

We reviewed the recent developments in
the type IIB matrix model, which was proposed as
a nonperturbative formulation of superstring theory 
in 1996.
While the Euclidean model has been shown to have
interesting dynamical properties, their physical interpretation
is yet to be clarified since the meaning of the Wick rotation
is not obvious.
On the other hand, the Lorentzian model remained untouched
until recently because of its instability, but recent Monte Carlo
studies revealed its surprising properties.
First of all, 
a well-defined theory can be obtained by introducing cutoffs
and removing them in the large-$N$ limit.
The notion of ``time evolution'' emerges dynamically.
This is due to the nontrivial dynamical property of the model
that the spatial matrices $A_i$ have a band-diagonal structure
when we diagonalize the temporal matrix $A_0$.
The extracted time evolution shows that, after some ``critical time'',
the space undergoes the SSB of SO(9) symmetry and only three
directions start to expand exponentially.
The observed exponential expansion suggests the possibility
that the Inflation is naturally realized in this model.
(Note that we do not introduce a scalar field
by hand, nor do we have to impose any particular initial condition.) 
We also observed the power-law $t^{1/2}$ expansion in
a simplified model for later times, which is reminiscent of the
cosmic expansion of the FRW universe in the radiation dominated era.
The behaviors at much later times are expected to be captured
by the classical equations of motion.
We have discussed a solution, which suggests a natural 
solution to the cosmological constant problem.

It would be very important to observe directly the transition
from the exponential expansion to the power-law expansion
by Monte Carlo simulation.
We speculate that 
the transition to commutative space-time
(as opposed to noncommutative one that is realized generically
in the matrix model)
occurs at the same time.
It would also be interesting to calculate the density fluctuation
to be compared with the cosmic microwave background.
Another direction would be to read off the effective 
quantum field theory below the Planck scale
from fluctuations around a classical solution that
dominates at later times \cite{Nishimura:2012rs}.
Along that direction, we should be able to see whether the 
Standard Model appears at low 
energy \cite{Chatzistavrakidis:2011gs,Aoki:2014cya,Steinacker:2014fja}.

To conclude, we would like to list fundamental
questions in particle physics and cosmology:
the mechanism of Inflation, the initial condition problem,
the cosmological constant problem,
the hierarchy problem,
dark matter, dark energy, baryogenesis,
the origin of the Higgs field,
the number of generations, etc..
It is conceivable that all these problems can be understood
in a unified manner by a nonperturbative formulation
of superstring theory.
The recent developments reviewed above seem
to suggest that the type IIB
matrix model indeed have the potential for such a formulation.

\begin{acknowledgement}
The author would like to thank the organizers of the
``Workshop on Noncommutative Field Theory and Gravity''
held in Corfu, Greece for hospitality.
He is also grateful to A.~Chatzistavrakidis and H.~Steinacker
for valuable discussions during the workshop.
This work is supported in part
by Grant-in-Aid for Scientific Research
(No.\ 20540286 and 23244057)
from JSPS.
\end{acknowledgement}

%
%

\end{document}